# An alternative application of GaAs-based light-emitting diodes: X-ray detection and imaging

Quan Yu[1], Fangbao Wang[2], Xin Yuan[1], Ying Liu[1], Lianghua Gan[3], Gangyi Xu[3,4], Wenzhong Shen[1,\*], Liang Chen[2,\*], Yueheng Zhang[1,\*]

1. *Key Laboratory of Artificial Structures and Quantum Control, School of Physics and Astronomy, Shanghai Jiao Tong University, Shanghai 200240, China*

2. *Radiation Detection Research Center, Northwest Institute of Nuclear Technology, Xi'an 710024, People's Republic of China*

3. *Key Laboratory of Infrared Imaging Materials and Detectors, Shanghai Institute of Technical Physics, Chinese Academy of Sciences, Shanghai 200083, China*

4. *Hangzhou Institute for Advanced Study, University of Chinese Academy of Sciences, Hangzhou 310024, China*

[\*]*Corresponding authors, E-mail addresses: wzshen@sjtu.edu.cn, Chenl_nint@163.com and yuehzhang@sjtu.edu.cn*

## Abstract

GaAs-based light-emitting diodes (LEDs) are commonly employed in a variety of applications, including medical imaging, biosensing, optical communications, and night vision. In this paper, we present an alternative application of GaAs-based LED with SI-GaAs substrate for X-ray detection and imaging. The mechanism relies on the semiconductor frequency down-conversion process, where the SI-GaAs substrate acts as a photodetector (PD). Upon X-ray irradiation, the generated photocurrent by the SI-GaAs substrate drives the LED to emit NIR photons which can be detect by a low-cost CCD. We demonstrate direct X-ray detection and present preliminary imaging results, providing another example of the applicability of the PD-LED design for optical frequency conversion. The proposed LED X-ray detector leverages mature materials and fabrication processes. The application of the frequency down-conversion concept makes it possible for pixel-less imaging using a large single imaging unit, eliminating the need for readout circuits. This PD-LED architecture offers an alternative approach to direct X-ray detection and imaging, characterized by higher absorption, improved image resolution, and enhanced material stability.

# Introduction

As one of the most important light emitting devices, light-emitting diodes (LEDs) are widely used in medical imaging[1], biosensing[2], optical communications[3,4], and night vision[5]. In the near-infrared (NIR) band, GaAs-based LEDs are of significant importance, attributed to their well-established material and processing technologies, simple structure and low price. GaAs-based LEDs can be grown on GaAs substrates of different doping type, including n-type, p-type[6] as well as semi-insulating GaAs (SI-GaAs) substrates. It is noted that SI-GaAs is an excellent material for X-ray detection[7] because of high absorption[8,9].

Since GaAs-based infrared LEDs can be epitaxially grown on SI-GaAs substrates, and the light emitted by the LEDs can be detected by a low-cost CCD directly, it naturally leads us to consider whether LEDs with SI-GaAs substrates can be directly used for X-ray detection and imaging. X-rays ionize and produce photogenerated charge carriers in the SI-GaAs. Under the influence of the electric field, these carriers move directionally to form photocurrent, which drives the LED to emit NIR photons. The emission can be directly observed using an infrared camera, thus achieving X-ray detection/imaging. This actually utilizes a frequency conversion detection and imaging mechanism, which has been successfully applied in the infrared band with several examples[10–16]. By integrating different type of photodetector (PD) with a LED, infrared as well as terahertz imaging[17] were realized. Considering the possibility of using SI-GaAs substrate as the PD for X-ray detection, a frequency down-conversion process can be realized by converting X-rays to NIR photons. It is not difficult to find that such kind of imaging scheme shows an obvious advantage since it does not need to connect readout circuits and has the potential to achieve pixel-less imaging using a large single imaging unit, therefore simplifying the manufacturing complexity. In contrast, in traditional GaAs X-ray direct imaging detectors needed to be integrated with readout circuits[18,19] to achieve efficient and high-resolution X-ray imaging. The proposed scheme is similar to conventional scintillation detectors[20,21] in that it belongs to the category of indirect imaging schemes. This kind of detection/imaging scheme fully leverages the advantages of GaAs X-rays detector and frequency conversion, which makes high absorption and high-resolution imaging possible.

In this paper, we introduce an alternative detection/imaging method that differs completely from traditional X-ray detection/imaging schemes. Utilizing the mechanism of frequency down-conversion, a simple infrared LED with a SI-GaAs substrate is employed as an X-ray detection/imaging device. Preliminary imaging results are obtained by etching through the heavily doped bottom contact layer. This proposed X-ray detection configuration is not only simple but also capitalizes on the properties of SI-GaAs, a commercially accessible X-ray detection material known for its high absorption and spatial resolution capabilities. This study marks the inaugural application of PD-LED architecture for frequency down-conversion imaging, demonstrating the wide applicability of the PD-LED design for optical frequency conversion.

## I. Device Structure and operation mechanism

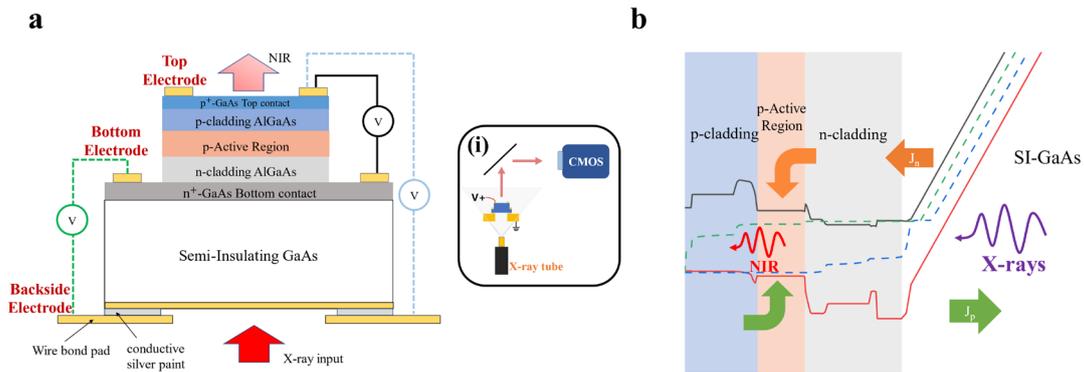

**Fig. 1 a** Schematic of the device structure, (i) inset is experimental setup for X-rays detection/imaging. **b** Band diagram and mechanism of the frequency down-conversion process.

The schematic structure of the device is shown in Fig. 1a. This LED material is epitaxially grown on a 625 μm SI-GaAs substrate in sequence: the highly doped $n^+$-GaAs bottom contact layer, the n-AlGaAs cladding layer, the p-active region, the p-AlGaAs cladding layer, and the highly doped $p^+$-GaAs top contact layer. By the standard photolithography and wet etching processes, square mesas with sizes of 1 mm ×1 mm and 3 mm×3 mm were fabricated. The mesa is etched to the bottom contact layer. We deposited Ti/Pt/Au on the p-type doped top contact layer and AuGe/Ti/Pt/Au on the n-type doped bottom contact layer to ensure ohmic contact. Ti/Pt/Au was deposited on the backside of SI-GaAs substrate to form a Schottky backside electrode. The backside of the device was glued onto a PCB holder with conductive silver paint, which has a window to allow X-rays to be incident on the device from the backside. Detailed parameters of LED are shown in Fig. S1. This epitaxial structure refers to the GaAs/AlGaAs LED described by D. Ban et al.[22] The active region is composed of 400 nm Be doped GaAs with the doping concentration of $1.0 \times 10^{18} cm^{-3}$, which is sandwiched between n-type and p-type cladding layers.

Considering the presence of three electrodes—namely the top, bottom, and backside electrodes—there are three distinct testing modes designed to measure the characteristics of the SI-GaAs photodetector (PD), the LED, and the X-ray down-conversion process, respectively, as depicted in Figure 1a. To solely measure the response of the SI-GaAs to X-rays, a bias was applied between the bottom and backside electrodes, with the bottom electrode being positive. When evaluating the LED's performance, a bias was applied across the top and bottom electrodes, with the top electrode being positive. This setup ensures that the LED is under forward bias, allowing us to examine its electroluminescent properties. For the X-ray down-conversion measurement, we applied a bias between the top electrode and the backside electrode, with the top electrode being positive to forward-bias the LED.

The schematic experimental setup is shown in the inset of Fig. 1a. The entire setup is placed in a darkroom made from lead-lined panels. X-ray is generated by an X-ray tube 1.5 cm away from the sample. The intensity of X-ray is controlled by the tube voltage $U_{tube}$ and tube current $I_{tube}$, which is proportional to $I_{tube}$, approximately quadratically related to $U_{tube}$, and inversely proportional to the square of the distance. With a tube voltage of 30 kV and a tube current of 100 μA, the calibrated radiation dose rate at a distance of 2 cm from the head of the X-ray tube is 0.383 y ·s$^{-1}$. Through the backside window of the PCB holder, the X-rays irradiate on the SI-GaAs substrate. The emitting light from LED is reflected by a 45° mirror and observed on the other side with the IR camera.

The operation mechanism of the device can be understood in terms of the energy band diagram shown in Fig. 1b. When the device operates in the X-ray down-conversion mode, the voltage mainly drops across the SI-GaAs region, since the LED is forward-biased. Upon X-ray irradiation, a large number of electron-hole pairs were generated in the SI-GaAs substrate. Under the electric field formed by the bias voltage, these carriers move directionally, i.e. electrons move towards the LED region and holes move in the opposite direction, forming both electron and hole currents. Electrons injected into the active region of the LED recombine radiatively with holes injected from the other side, thus emitting NIR photons. The transformation of X-ray photons to NIR photons within the device is a frequency down-conversion process.

## II. X-ray detection

To investigate the performance of the down-conversion X-ray detector, understanding the characteristics of SI-GaAs substrate as a PD and the LED is necessary. In the following, we first measure the performance of the SI-GaAs substrate and the LED. The size of the tested device is $1\text{mm} \times 1\text{mm}$.

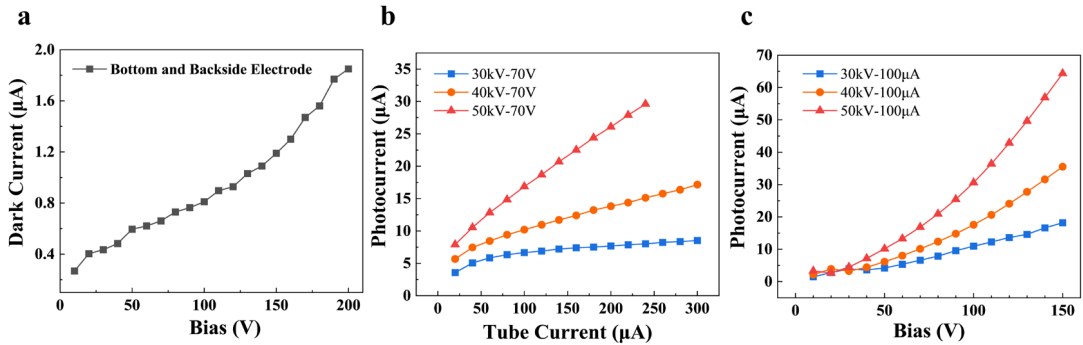

**Fig. 2** The dark current and photocurrent response. **a** The dark current characteristics of the SI-GaAs substrate as a PD. **b** The dependence of photocurrent of the SI-GaAs substrate on the X-ray tube current, when $U_{bias} = 70V$, and $U_{tube} = 30kV, 40kV, 50kV$. **c** The dependence of photocurrent of the SI-GaAs substrate on the bias under three different X-ray intensities: $(U_{tube} = 30kV, I_{tube} = 100\mu A)$, $(U_{tube} = 40kV, I_{tube} = 100\mu A)$, $(U_{tube} = 50kV, I_{tube} = 100\mu A)$.

Fig. 2a presents the dark current characteristics of SI-GaAs substrate as a PD, where the bottom electrode and the backside electrode are biased. The SI-GaAs substrate exhibited high resistance without X-ray irradiation. The dark current is below 2 µA, even when the applied voltage is up to 200 V. Under X-ray irradiation, photocurrent is generated, which could be influenced by the tube current $I_{tube}$ and tube voltage $U_{tube}$ as well as the bias. Fig. 2b shows the linear dependence of the photocurrent on the tube current under different tube voltage when the bias of the device is 70V. The slight deviation from linearity at tube current less than 50 µA may be attributed to the X-ray tube's tungsten filament not reaching a high enough temperature when operating at low current. Because the intensity of X-rays is proportional to the tube current, and the generated photocurrent is also proportional to the X-ray intensity, the linearity observed in the graph is reasonable. It is noted that the higher the tube voltage, the stronger the generated photocurrent. Additionally, the photocurrent varies with the bias voltage of the device. As shown in Fig. 2c, three different X-ray intensities were used: ($U_{tube} = 30\text{kV}, I_{tube} = 100\text{µA}$), ($U_{tube} = 40\text{kV}, I_{tube} = 100\text{µA}$), and ($U_{tube} = 50\text{kV}, I_{tube} = 100\text{µA}$). With increasing bias, the electric field across the SI-GaAs becomes stronger, leading to higher collection efficiency of the carriers generated by the X-rays, and consequently a larger photocurrent. It should be noted that due to the very high resistance of SI-GaAs substrate, the dark current of the SI-GaAs substrate PD is almost negligible compared to the photocurrent, which indicates that light emission of the LED is mainly driven by the photocurrent.

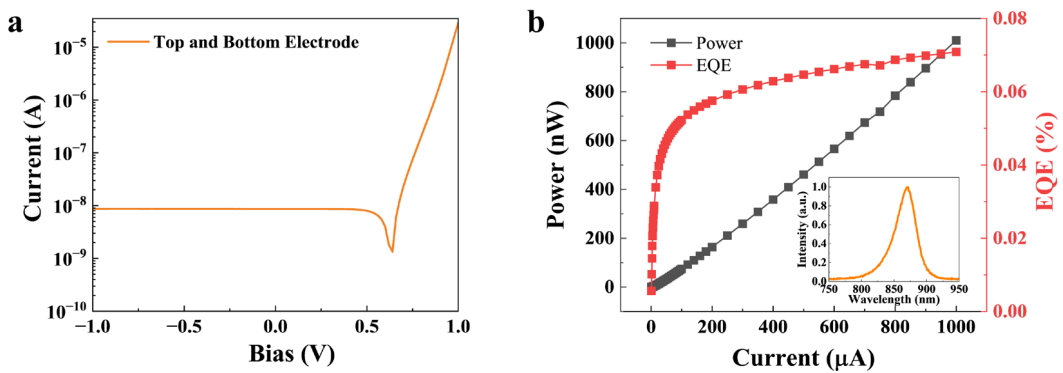

**Fig. 3 a** I-V characteristics of the LED, biasing the top and bottom electrodes, and **b** external quantum efficiency (EQE) and optical power versus current of the LED. The inset is emission spectrum of the LED.

To study the properties of LED, we apply a bias between the top and bottom electrodes. Fig. 3a shows the I-V curve of the LED alone. It clearly exhibits a rectification character, indicating that the LED structure can be simply seen as a p-n junction diode. Fig. 3b demonstrates the dependence of the optical power as well as the external quantum efficiency (EQE) on the driving current. It can be seen that the light emission power is directly proportional to the driving current, with the emission peak around 870 nm. The external quantum efficiency is a key parameter to characterize the performance of LED, which is defined as the ratio of the number of the outgoing photon to that of the electron injected to LED. Considering the bias, the tube voltage and tube current for device operation, the photocurrent generated by SI-GaAs is about tens of µA, leading to the light output power is about 10~70 nW, and the EQE is about 0.05 %. The low EQE is due to the device operating at a low driving current, where the injection efficiency of the LED is not high, dominated by SRH recombination processes[22,23]. Though the EQE is not high, it is sufficient to observe the frequency down-conversion process.

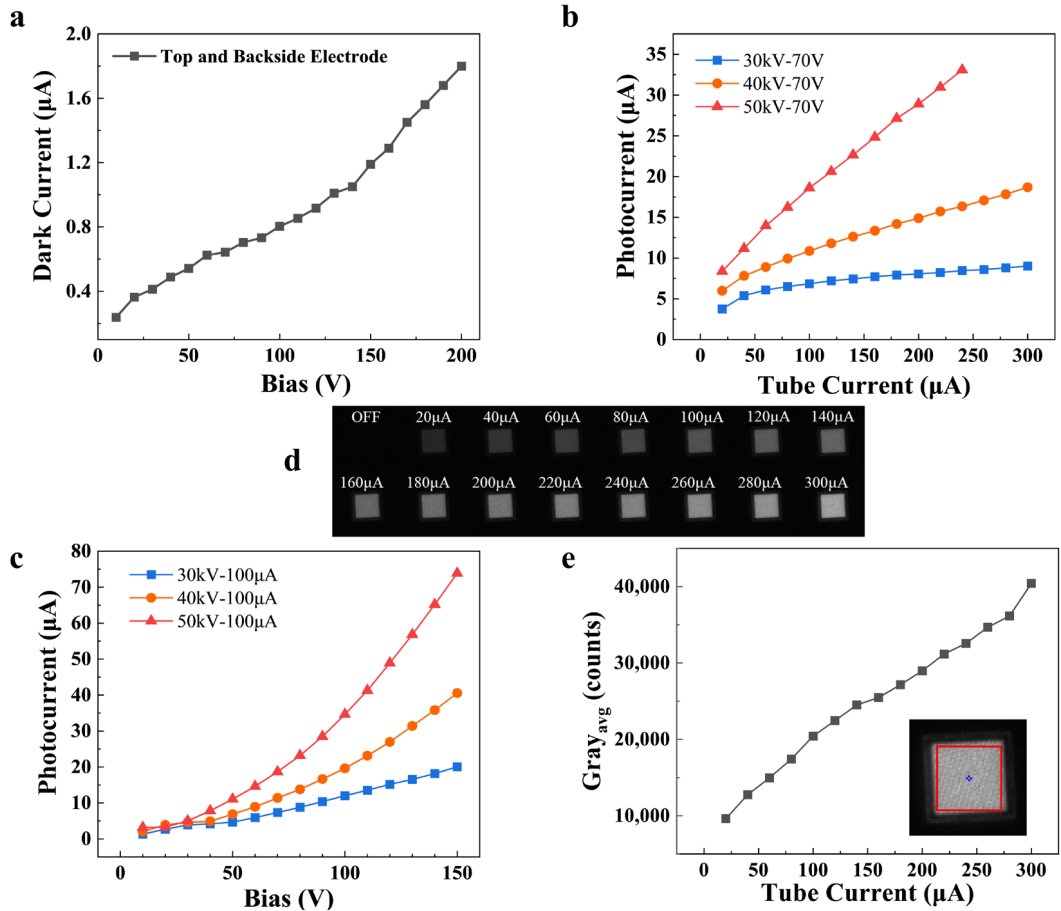

**Fig. 4 a** I-V characteristics of the device working in the down-conversion mode. **b** Photocurrent response to X-ray tube current at $U_{bias} = 70V$, when $U_{tube}$ is $30kV, 40kV, 50kV$ respectively. **c** The dependence of photocurrent on the bias, when $I_{tube}$ is $100\mu A$ and $U_{tube}$ is $30kV, 40kV, 50kV$ respectively. **d** Infrared photographs of the luminescence produced by a 1mm device under X-ray irradiation with different $I_{tube}$, when $U_{bias} = 70V$ and $U_{tube} = 40kV$. **e** The average grayscale values of the above infrared photographs under different $I_{tube}$.

Based on the investigation of the SI-GaAs substrate (as a PD) and of the LED, we further studied the characteristics of the device as a frequency down-conversion X-ray detector by applying the bias between the top and the backside electrode. Fig. 4 shows the dark current as well as the photocurrent response to X-rays. It is noted that the characteristics and the magnitude of the dark current and the photocurrent are almost identical to that of SI-GaAs substrate. This indicates that the SI-GaAs substrate is the main region where photocurrent is generated and the voltage drops. The SI-GaAs substrate of LED can indeed function as a PD, which generates photocurrent and drives the LED.

Through the frequency down-conversion process, the X-ray detection was realized by capturing the luminescence of the LED using an infrared camera, as shown in Fig. 4d. This figure displays a set of images of a 1 mm-sized mesa under X-ray irradiation of different tube currents (20 μA to 300 μA), when the bias voltage is 70 V and $U_{tube}$ is 40 kV. The image of the device without X-rays irradiation is also presented and no LED emission can be observed through the camera because of the low dark current (<2 μA). The integration time of the camera is 100 ms. It can be observed that the infrared luminescence of the down-conversion device becomes stronger as the X-ray intensity increases. The minimum emitted optical power of the LED that can be observed by the infrared camera is about 0.6 nW, which requires the driving current to be at least 3 μA. It indicates that without X-ray irradiation, no LED emission can be observed through the camera because of the low dark current (<2 μA). The average grayscale value ($ray_{avg}$) of the above infrared photos is also examined as shown in Fig. 4e. The value is calculated by averaging the grayscale value in the 1mm area (indicated in red) of the infrared photo. It can be seen that $ray_{avg}$ varies linearly with tube current as the photocurrent does.

So far, we have demonstrated that X-ray detection could indeed be realized with a simple LED by the frequency down-conversion mechanism. Whether this device can be used for imaging directly is still an open question. In the following, the capability for X-ray spatial resolution was studied.

## III. X-ray imaging

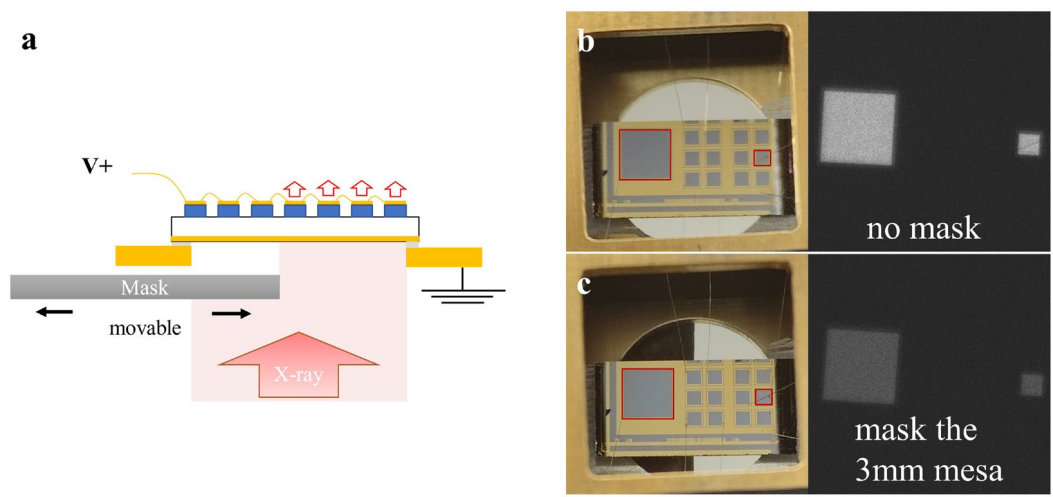

**Fig. 5 a** Schematic experimental setup for spatial resolution testing. All the mesas share the same bottom contact layer. **b** The actual image (left) and corresponding IR image (right) of the device with no mask. The two parallel-connected mesas (one 3 mm mesa and one 1 mm mesa) are marked in red. **c** The actual image (left) and IR image (right) after masking the 3 mm mesa.

To verify if such down-conversion X-ray detector can be used for imaging directly, the capability for X-ray spatial resolution is investigated. The schematic diagram of the experiment setup is shown in Fig. 5a. The mesas are connected in parallel using gold wires. A movable mask is utilized to block the X-rays incident on the sample, resulting in some mesas not being irradiated by X-rays, while others are. We expect that only the irradiated mesas would emit light if it has spatial resolution. Based on such consideration, we fabricate multiple mesas on the same epitaxial wafer. We connected two mesas with sizes of 3 mm and 1 mm in parallel, as indicated in red in Fig. 5b and Fig. 5c. The down-conversion X-ray-to-infrared luminescence without a mask is shown in Fig. 5b, where both mesas glow simultaneously. In contrast, the down-conversion infrared photograph of the 3mm mesa blocked from X-rays is presented in Fig. 5c. Contrary to our expectation, even though the 3 mm mesa was masked, both the 3 mm and 1 mm mesas glow simultaneously regardless of whether they were blocked or not. This indicates that such a structure lost the spatial resolution of imaging. The current driving the 3 mm mesa was laterally conducted from the unmasked portion. Due to the masking, the total photocurrent generated was smaller, hence the infrared luminous intensity was also weaker compared to the no mask scenario.

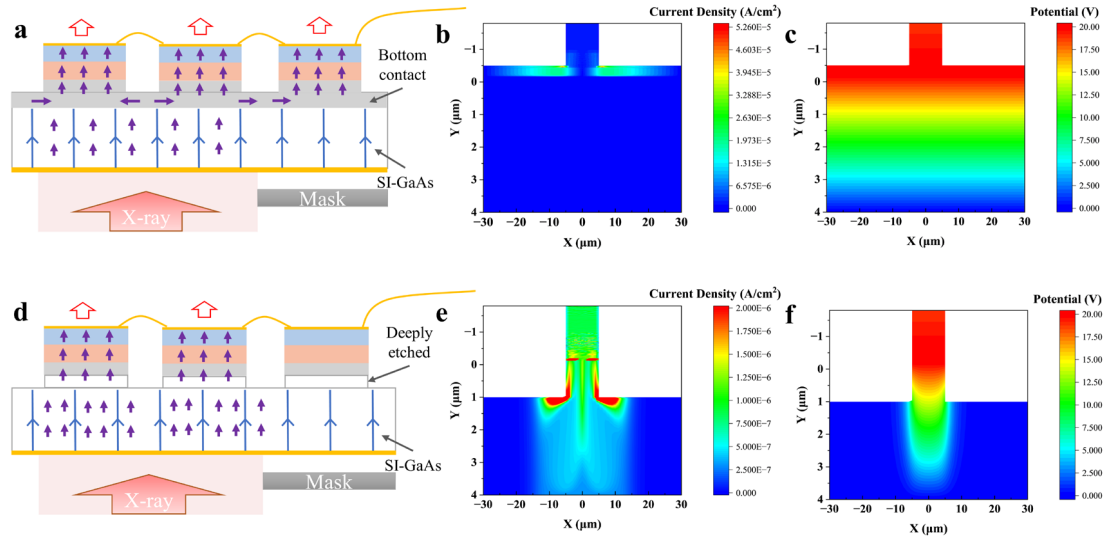

**Fig. 6** Simulation of the effect of the bottom contact layer on lateral current conduction. **a** Schematic diagram of the photocurrent distribution in the device retaining the bottom contact layer. **b** The current density in the device retaining the bottom contact layer. **c** The electric potential in the device retaining the bottom contact layer. **d** Schematic diagram of the photocurrent distribution in the device with the bottom contact layer etched away. **e** The current density in the device with the bottom contact layer etched away. **f** The electric potential in the device with the bottom contact layer etched away.

To understand why the device lost the capability of spatial resolution, we inspect the epitaxial structure and fabrication process. It is noted that all the mesas share the common bottom contact layer. Due to the high doping and low resistance of this layer, the photocurrent will be easily conducted laterally in the bottom contact layer. Fig. 6a shows the schematic photocurrent distribution in the device, when multiple mesas are connected in parallel and the device is partially masked from X-rays. Due to the high conductivity of the bottom contact layer, the electric field across the SI-GaAs region is uniform when biased. Upon irradiation with X-rays, a large number of photogenerated carriers are produced in the unmasked area of the SI-GaAs. These carriers drift under the uniform electric field and reach the bottom contact layer. So far, the spatial information has not yet been lost. Once the carriers reach the bottom contact layer, due to the high conductivity of the p-doped contact layer, the photocurrent will conduct laterally leading to a redistribution of the current, with the final effect of making all the parallel-connected mesas glow simultaneously. To verify this assumption, we simulate the current distribution of a single mesa, and there is an obvious lateral spreading of the current in the bottom contact layer, as shown in Fig. 6b. The potential distribution is shown in Fig. 6c, which is laterally uniform in this geometry. It indicates that a uniform electric field is distributed in the SI-GaAs. This structure is analogous to a planar capacitor, with the highly doped layer and the backside metal serving as the two metal plates, and the SI-GaAs in between acting as the dielectric layer.

The lateral conduction effect of the bottom contact layer can also be inferred from the magnitude of the photocurrent. We measure the photocurrent of 1 mm mesa and 3 mm mesa using the X-ray with the same intensity, respectively. The photocurrent is almost equal in magnitude. The results are shown in Fig. S2 of Supplementary Material. The resulted photocurrent is even the same as that measured by the SI-GaAs substrate (biasing the bottom electrode and backside electrode). If the device has the spatial resolution, the photocurrent collected by the 3 mm size mesa would be larger than that of the 1 mm size, but this is not the case. This is actually due to the fact that the presence of the bottom contact layer collects the photocurrent generated by the entire SI-GaAs substrate and injects it into a single mesa, therefore, the magnitude of current is independent of the size of that mesa.

In order to eliminate the impact of the bottom contact layer on the current conduction, we increased the etch depth of the mesas, etching through the highly conductive common bottom contact layer and revealing the highly resistive SI-GaAs. The modified structure is shown in Fig. 6d. The etching depth of all the mesas exceeds the thickness of the epitaxial layer, and reaches the SI-GaAs region. In this deeply-etched structure, the mesas are able to collect the photocurrent at the corresponding position without significant current spreading. Fig. 6e shows the corresponding simulation results. It can be seen that that the current distribution is more concentrated under the mesas, and the lateral current spreading is significantly suppressed. In Fig. 6f, the simulated potential drops mainly in the region below the mesa, which indicates that the electric field in this region is more concentrated, and this localization enables the mesa to collect the photocurrent at the corresponding position.

In contrast, the collected photocurrent in the deeply-etched structure is measured. Fig. S3 in Supplementary Material gives the magnitudes of collected photocurrent of one 3 mm mesa and one 1 mm mesa. It is seen that the magnitude of the photocurrent is now correlated with the area of the mesa. Additionally, the photocurrent is lower than that of the structure with the common bottom contact layer. These results indicate that the deeply-etched structure no longer collects photocurrent generated at other positions, and current spreading is effectively suppressed.

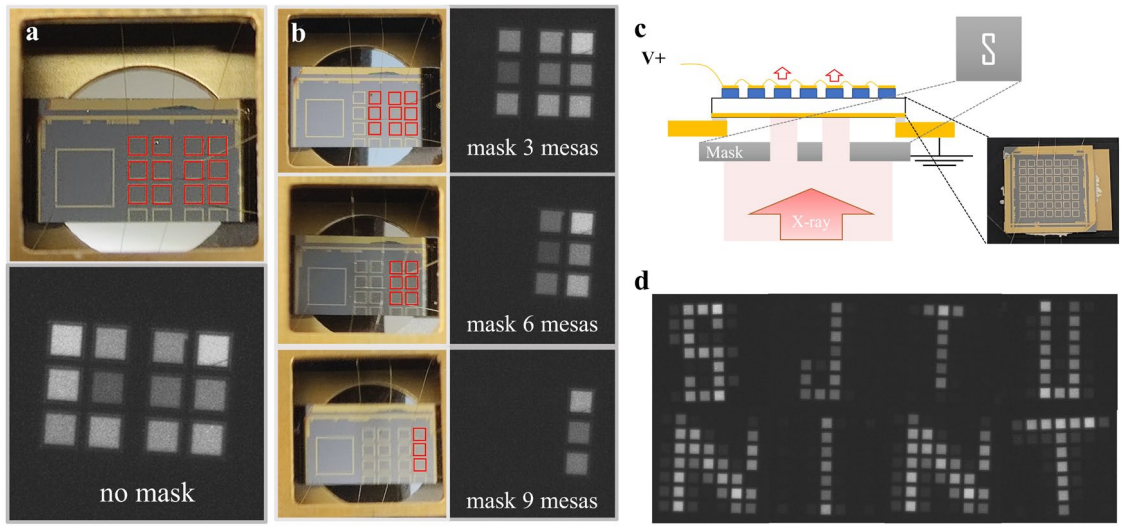

**Fig. 7** Spatial resolution results for the structure with the bottom contact layer etched away, showing the actual devices and corresponding down-conversion infrared photographs. **a** The actual device image and IR photograph, with 12 parallel-connected 1mm mesas. **b** Different numbers of masked mesas are shown, masking 3, 6, and 9 mesas from top to bottom, respectively. **c** A 7×7 array device for down-converted imaging, with each mesa size of $1 \text{ mm}^2$, using a letter-shaped mask to block the incident X-rays. **d** Infrared photographs of down-converted imaging after blocking X-rays with different letter-shaped masks.

We performed X-ray spatial resolution tests on a deeply-etched device. 12 mesas of 1 mm size in parallel with gold wires are connected, as marked in red in Fig. 7a. The down-converted luminescence of these 12 mesas irradiated by X-ray without a mask is shown in the infrared photo below. Using a mask to cover different portions of the device, we obtained infrared photographs with different numbers of luminous mesas, see Fig. 7b. From top to bottom, 3, 6, and 9 mesas were masked, and the IR photos show the corresponding luminance. This shows that etching through the common bottom contact layer can greatly suppress the lateral current conduction and restore the spatial resolution capability of X-rays. It should be noted that the total thickness of the device is ~600μm. Compared with the case of the QWIP-LED[12,13] with the thickness of ~10μm, the thickness of our down-conversion X-ray detector is much thicker. It indicates that the spatial information of the light field could be retained well at this thickness as long as the resistance is high enough, proving the ubiquity of the frequency conversion imaging scheme.

To further investigate the imaging ability of the device, we fabricated an array-like device with more mesas, each of which is 1 mm in size, forming a 7×7 array, and the top electrodes of each mesa were connected in parallel via gold wires, enabling all mesas to be biased simultaneously. The actual shape of the device is shown in the inset in Fig. 7c. A mask with an alphabetic shape is placed at the back of the device to block the X-ray. The down-converted luminescence of the device also shows the shape of the corresponding letter, as seen in Fig. 7d. Here, we used down-converted imaging photos of different alphabetic characters to compose this figure.

In Fig. 7d, we can observe a distinct luminescence from the mesa irradiated by X-ray. However, it is also noted that there are some lightly glowing mesas at the edges of the letters, indicating that there is still some electric crosstalk. When collecting the photocurrent, some current flows into the mesa not irradiated by X-rays, resulting in a weak glowing. Although we have removed the bottom contact layer, which is the main conduction layer, the semi-insulating substrate can still laterally conduct the current to some extent, as shown in Fig. 6e where the current collecting area is larger than the mesa. This diffusion effect may be suppressed by using a SI-GaAs substrate with a higher resistivity and thinner thickness or by lowering the operating temperature of the device. Additionally, the device has some malfunctional mesas that won't glow at all, and the brightness between the mesas is not uniform. This may be related to defects, crystal growth quality, and the sophistication of the manufacturing process. These challenges encountered in up-conversion pixel-less imaging device either, which need large areas of material without severe defects. However, in the case of such an array imaging, the defects only render individual mesas inoperable while the rest can still operate normally, but in infrared up-conversion pixel-less imaging, there is only one large imaging unit, severe defects would cause the entire device to fail. Therefore, this array-like pixelated PD-LED configuration is more tolerant to defects. Furthermore, the current mesa area demonstrated is 1 mm$^2$. If we increase the number of pixels and make them denser, a higher imaging resolution could be realized.

As originally conceived, such frequency conversion devices can be used for pixel-less imaging similar to up-conversion results of QWIP-LED[12,13]. So, we performed detailed tests on a 3 mm mesa of a deeply-etched structure (see Fig. S4). Unfortunately, we failed to achieve pixel-less imaging. Within the large-sized mesa the luminance was uniform and the spatial information was lost. The highly doped layer in the mesa again acted as a current spreading layer. Another possible reasons why we failed to achieve pixel-less imaging is that our device operates at room temperature and carriers have a large diffusion coefficient, whereas QWIP-LEDs operate at liquid nitrogen temperature. By cooling down the device we may restore the spatial information inside the mesa and achieve pixel-less imaging. Alternatively, the common region of the LED and SI-GaAs could be improved, for instance, by fabricating an i-n-i common region[14,24]. By reducing the doping concentration of the bottom contact layer to suppress the lateral current conduction, it may also be possible to achieve pixel-less imaging.

## IV. Discussion

As one of the most important light emitting device in near infrared band, GaAs-based LED shows high performance, well-established manufacturing processes and reduced production costs. In this paper, we have demonstrated an alternative application of a simple GaAs-based LED with SI-GaAs substrate for direct X-ray detection and imaging. The underlying mechanism involves semiconductor frequency down-conversion, where the SI-GaAs substrate serves as a detector. When exposed to X-rays,

SI-GaAs substrate generates photocurrent that drives the LED to emit NIR photons, which are subsequently detected using an infrared camera. This approach offers several advantages, including the elimination of the need for a readout circuit, thereby simplifying manufacturing complexity. The device exhibits a linear response to X-rays and operates efficiently at room temperature.

Furthermore, we have explored the role of the heavily doped bottom contact layer of the LED in current conduction. By increasing the etching depth, we have significantly reduced its impact on imaging, achieving preliminary X-ray imaging results. This work provides another example of the applicability of the PD-LED design for optical frequency conversion. To improve the imaging results, we suggest fabricating devices with a higher pixel count to obtain an array-like detector. Alternatively, to fabricate a large-area pixel-less imaging detector, the connection region between the LED and the SI-GaAs should be specially designed. For instance, an intrinsic-n-intrinsic (i-n-i) common region would be feasible. Reducing the doping concentration and the thickness of the highly doped bottom contact layer could suppress lateral current conduction, potentially enabling pixel-less imaging.

Although this paper only present preliminary imaging results, the advantages of such a down-conversion X-ray detector in imaging are noteworthy, and its potential applications are promising. Firstly, it is based on the mature material and fabrication processing. As a commercial X-ray detection material, the SI-GaAs material itself offers high absorption and spatial resolution capabilities. Compared with traditional scintillators, such as CsI(Tl)[25], NaI(Tl)[26], $Bi_4e_3O_{12}$ (BGO)[27] and $(Lu,Y)SiO_5$[28], SI-GaAs not only exhibit excellent stability and optoelectrical properties, but also is non-toxic and environmentally friendly. Secondly, the frequency down-conversion concept of the PD-LED design used in this work can be flexibly extended to other material systems. For example, it can be used to achieve down-conversion from X-rays to visible light using other inorganic semiconductors like GaN[29,30], or by integrating organic OLEDs with SI-GaAs to create hybrid X-rays-to-visible-light down-conversion devices. Leveraging the low lateral diffusion characteristics of carriers in organic materials[31] and the advantages of solution-based[32,33] fabrication, it will be easier to produce large-area, pixel-less imaging devices. We believe that based on the PD-LED architecture, it is possible to fabricate X-ray imaging devices with higher absorption, improved image resolution, and enhanced material stability, paving the way for future advancements in medical and industrial imaging technologies.

## V. Methods

**Measurement details.** The power source meters used in the experiment were Keithley 2400 and KEISIGHT B2902A Precision power supply. The X-ray source was a Moxtek ULTRA-LITE MAGNUM X-ray tube. The IR camera was a Photonic Science cooled sCMOS Camera. The emission spectrum and optical power of the LED were measured using an Ocean optics QE65PRO fiber spectrometer and a Thorlabs S130C large area Si slim photodiode power meter, respectively.

## VI. Acknowledgements

This work was supported by the Natural Science Foundation of China (Grant Nos. 12274285，12305205，62204198, 12393833, 62435020, 62235010), the Shanghai New Energy Technology Research and Development Project, China (No. 24DZ3000900).

Supplementary Information for

# An alternative application of GaAs-based light-emitting diodes: X-ray detection and imaging

Quan Yu[1], Fangbao Wang[2], Xin Yuan[1], Ying Liu[1], Lianghua Gan[3], Gangyi Xu[3,4], Wenzhong Shen[1,] *, Liang Chen[2,] *, Yueheng Zhang[1,] *

1. *Key Laboratory of Artificial Structures and Quantum Control, School of Physics and Astronomy, Shanghai Jiao Tong University, Shanghai 200240, China*
2. *Radiation Detection Research Center, Northwest Institute of Nuclear Technology, Xi'an 710024, People's Republic of China*
3. *Key Laboratory of Infrared Imaging Materials and Detectors, Shanghai Institute of Technical Physics, Chinese Academy of Sciences, Shanghai 200083, China*
4. *Hangzhou Institute for Advanced Study, University of Chinese Academy of Sciences, Hangzhou 310024, China*

[*]Corresponding authors, E-mail addresses: wzshen@sjtu.edu.cn, Chenl_nint@163.com and yuehzhang@sjtu.edu.cn

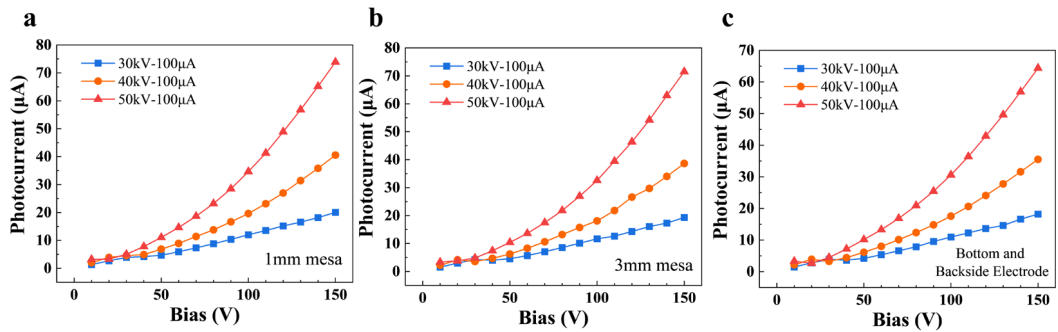

**Fig. S1** Detailed parameters of epitaxial LED. Refer to D. Ban *et al.*[1]

Detailed parameters of LED are shown in Fig. S1. This epitaxial structure refers to the GaAs/AlGaAs LED described by D. Ban *et al.*[1] This LED exhibits good room-temperature performance, emitting light peaked at the near-infrared wavelength of 870 nm. The active region is composed of 400 nm Be doped GaAs with the doping concentration of $1.0 \times 10^{18} \text{cm}^{-3}$, which is sandwiched between n-type and p-type cladding layers.

**Fig. S2** Comparison of photocurrent collection capability and mesa size in devices retaining the bottom contact layer. **a** Photocurrent versus bias voltage for a device with a mesa size of 1 mm. **b** Photocurrent versus bias voltage for a device with a mesa size of 3 mm. **c** Photocurrent versus bias voltage for SI-GaAs substrate (biasing the bottom and backside electrodes).

Fig. S2 shows the photocurrent generated by the structure retaining the bottom contact layer. We show the photocurrent response of the 1 mm mesa, the 3mm mesa and the SI-GaAs substrate respectively (Fig. S2 a-c). It can be observed that the magnitudes of the photocurrent to the same X-rays intensity are nearly identical, independent of the mesa size. This indicates the effect of the common bottom contact layer on current collection. This layer collects the photocurrent from the entire substrate and injects it into a single mesa, and thus the magnitude of photocurrent is independent of the size of that mesa.

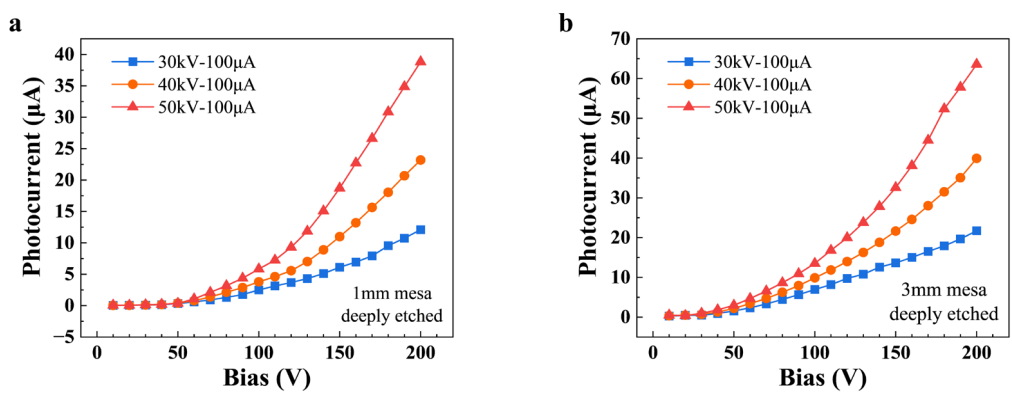

**Fig. S3** Comparison of photocurrent collection capability and mesa size in devices with the bottom contact layer etched away. **a** Photocurrent versus bias voltage for a device with a mesa size of 1 mm. **b** Photocurrent versus bias voltage for a device with a mesa size of 3 mm.

In Fig. S3, we compared the photocurrent response of mesas of different size on a deeply etched sample. The mesa sizes are 1 mm and 3 mm respectively. It was observed that the 3 mm mesa exhibited a stronger capability of photocurrent collection than the 1 mm one. Since the testing was conducted on a single mesa with other mesas on the same sample not being biased, the collected area of photocurrent could be larger than the tested mesa itself, resulting in the ratio of the photocurrent between the 1mm mesa and the 3mm one not equal to 1:9.

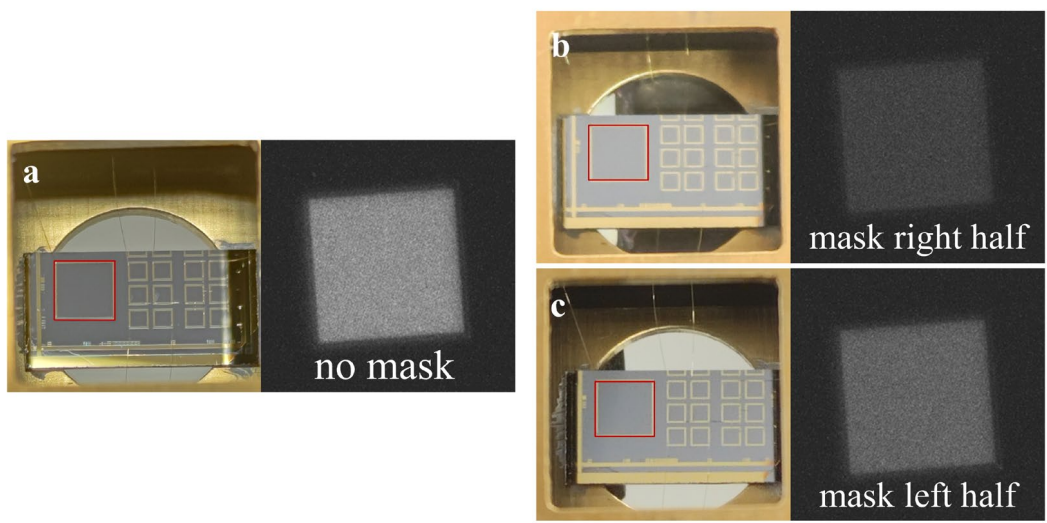

**Fig. S4** Spatial resolution results for a 3 mm mesa device with the bottom contact layer etched

away, showing the actual devices and corresponding down-conversion infrared photographs. **a** The luminance of the 3 mm mesa without mask. **b** Mask the right half of the 3 mm mesa. **c** Mask the left half of the 3 mm mesa.

The deeply etched structure, while capable of preserving spatial information between different mesas, still loses the spatial information within the mesa, as shown in Fig. S4. In Fig. S4a, it illustrates the light emission from a 3 mm mesa without masking, followed by masking the right half and left half of the mesa respectively (see Fig. S4 b, c). The infrared photos shows that the glowing of the mesa is uniform, with only a reduction in intensity due to the masking. This indicates that the spatial information is once again lost within the highly doped layer inside the mesa. For PD-LED structures operating at room temperature, a specially designed interlayer is needed to suppress the lateral current diffusion[2].